\documentclass[structabstract]{aa}  
\usepackage{graphicx}
\usepackage{txfonts}

\def\msun{M$_{\odot}$}
\def\h{$^{\rm h}$}\def\m{$^{\rm m}$}

\def\degs{\ifmmode ^{\circ}\else$^{\circ}$\fi}
\def\fss{\hbox{$.\!\!^{\rm s}$}}        
\def\fdg{\hbox{$.\!\!^\circ$}}          
\def\farcs{\hbox{$.\!\!^{\prime\prime}$}}  

\def\amin{\ifmmode ^{\prime}\else$^{\prime}$\fi}
\def\asec{\ifmmode ^{\prime\prime}\else$^{\prime\prime}$\fi}
\newbox\grsign \setbox\grsign=\hbox{$>$}
\newdimen\grdimen \grdimen=\ht\grsign
\newbox\laxbox \newbox\gaxbox
\setbox\gaxbox=\hbox{\raise.5ex\hbox{$>$}\llap
     {\lower.5ex\hbox{$\sim$}}}\ht1=\grdimen\dp1=0pt
\setbox\laxbox=\hbox{\raise.5ex\hbox{$<$}\llap
     {\lower.5ex\hbox{$\sim$}}}\ht2=\grdimen\dp2=0pt

\def\lax{$\mathrel{\copy\laxbox}$}
\begin{document}
   \title{The redshift and afterglow of the extremely energetic \\ gamma-ray
    burst GRB 080916C}

   \author{J. Greiner\inst{1} \and
           C. Clemens\inst{1} \and
           T. Kr\"uhler\inst{1,2} \and
           A. v. Kienlin\inst{1} \and
           A. Rau\inst{3} \and
           R. Sari\inst{4} \and
           D.B. Fox\inst{5} \and
           N. Kawai\inst{6} \and
           P. Afonso\inst{1} \and
           M. Ajello\inst{7} \and
           E. Berger\inst{8} \and
           S.B. Cenko\inst{9} \and
           A. Cucchiara\inst{5} \and
           R. Filgas\inst{1} \and
           S. Klose\inst{10} \and
           A. K\"{u}pc\"{u} Yolda\c{s}\inst{11} \and
           G.G. Lichti\inst{1} \and
           S. L\"ow\inst{1} \and
           S. McBreen\inst{12,1} \and
           T. Nagayama\inst{13} \and
           A. Rossi\inst{10} \and
           S. Sato\inst{14} \and
           G. Szokoly\inst{15,1} \and
           A. Yolda\c{s}\inst{1} \and
           X.-L. Zhang\inst{1}
          }

   \institute{Max-Planck-Institut f\"ur extraterrestrische Physik,
              Giessenbachstrasse 1, 85748 Garching, Germany\\
              \email{jcg@mpe.mpg.de}
         \and
            Universe Cluster, Technische Universit\"{a}t M\"{u}nchen,
           Boltzmannstra{\ss}e 2, D-85748, Garching, Germany 
         \and
            Optical Observatories, California Inst. of Technology,
              1200 E California Blvd, Pasadena, CA 91125, USA\\
            \email{arne@astro.caltech.edu}
          \and
            Dept. of Theor. Astrophysics, California Inst. of Technology, 
              1200 East California Blvd., Pasadena, CA 91125, USA \\
            \email{sari@tapir.caltech.edu}
         \and
             Department of Astronomy \& Astrophysics,
              Pennsylvania State University, 525 Davey Lab,
              University Park, PA 16802, USA \\
            \email{dfox@astro.psu.edu}
         \and
             Dept. of Physics, Tokyo Inst. of Technology,
              2-12-1 Ookayama, Meguro-ku, Tokyo 152-8551, Japan\\
             \email{nkawai@phys.titech.ac.jp}
         \and
              SLAC/KIPAC, 2575 Sand Hill Road, Menlo Park, CA 94025, USA \\
             \email{majello@slac.stanford.edu}
         \and
             Harvard University, 60 Garden Street, Cambridge, MA 02138, USA \\
              \email{eberger@cfa.harvard.edu}
         \and
              Department of Astronomy, University of California, Berkeley, 
              CA 94720, USA\\
              \email{cenko@astro.berkeley.edu}
         \and
             Th\"uringer Landessternwarte Tautenburg, Sternwarte 5,
              D-07778 Tautenburg,  Germany \\
            \email{klose@tls-tautenburg.de}
         \and
             ESO, Karl-Schwarzschild-Str. 2, 85740 Garching, Germany \\
             \email{ayoldas@eso.org}
         \and
             School of Physics, University College Dublin, Belfield, Dublin 4,
             Ireland \\
             \email{sheila.mcbreen@ucd.ie}
         \and
             Dept. of Astronomy, Kyoto University, Sakyo-ku, Kyoto 606-8502, 
              Japan \\
             \email{nagayama@kusastro.kyoto-u.ac.jp}
         \and
              Department of Astrophysics, Nagoya University, Furo-cho,
             Chikusa-ku, Nagoya 464-8602, Japan \\
             \email{ssato@z.phys.nagoya-u.ac.jp}
         \and
            E\"otv\"os Univ., 1117 Budapest, Pazmany P. stny. 1/A, Hungary \\
             \email{szgyula@elte.hu}
             }

   \date{Received 22 Dec 2008; accepted 2009}

 
  \abstract
   {The detection  of GeV photons from gamma-ray  bursts (GRBs) has
    important  consequences for  the interpretation  and  modelling of
    these   most-energetic    cosmological   explosions.    The   full
    exploitation of  the high-energy measurements  relies, however, on
    the accurate  knowledge of the  distance to the events.}
    {Here we report on the discovery of the afterglow  and subsequent redshift
    determination of GRB~080916C, the
    first GRB detected by the Fermi Gamma-Ray Space Telescope with 
     high significance detection of photons at energies $>$0.1 GeV.
    }
   {Observations were done with 7-channel imager GROND at the 
    2.2m MPI/ESO telescope,
    the SIRIUS instrument at the Nagoya-SAAO 1.4\,m telescope in South Africa, 
    and the GMOS instrument at Gemini-S. 
   }
    {The afterglow photometric
    redshift  of  $z=4.35\pm0.15$,   based  on  simultaneous  7-filter
    observations with the Gamma-Ray Optical and Near-infrared Detector
    (GROND), places GRB~080916C among the top 5\,\% most
    distant GRBs,  and makes it the most energetic
    GRB known to date. The detection of GeV photons from such
    a distant event is rather  surprising.
    The observed  gamma-ray
    variability  in the  prompt  emission together  with the  redshift
    suggests   a  lower   limit  for   the  Lorentz   factor   of  the
    ultra-relativistic ejecta  of $\Gamma > 1090$.   This value rivals
    any previous measurements of  $\Gamma$ in GRBs and strengthens the
    extreme nature of GRB~080916C.
   }
   {}


   \maketitle
%

\section{Introduction}

\begin{table*}
  \caption{Log of the observations\label{log}.
    The first ground-based imaging was obtained 
    with the Simultaneous 3-color ($JHK$) InfraRed Imager for Unbiased
    Survey (SIRIUS, \cite{nag03})
    on the Nagoya-SAAO 1.4m telescope (IRSF).
    GROND, a simultaneous 7-channel imager \cite{gbc08} 
    mounted at  the 2.2\,m
    MPI/ESO telescope at La Silla (Chile), started observing about 30.75\,hrs after the GRB.  The imaging sequence consisted
    of a series of sixteen 375\,s integrations in the $g'r'i'z'$ channels with 
    gaps of about 45\,s.
    In parallel, the  $JHK_S$ channels were operated with
    10\,s integrations, separated by 5\,s.
    Late-time imaging was obtained with the Gemini-South telescope +
    Gemini Multi-Object Spectrograph (GMOS-South) on 29 Oct 2008, taking
    eight 180\,s exposures.
    Data reduction was done using IRAF routines.
    Photometric calibration of the GROND $g',r',i',z'$ bands was performed
    using the spectrophotometric standard stars SA100-241 and SA97-249,
    while that of $JHK_S$ was done against 2MASS (Tab. \ref{stand}).
  }
\small
\begin{tabular}{llccc}
   \noalign{\smallskip}
   \hline
   \noalign{\smallskip}
   ~~~~Date/Time  & Telescope/Instrument & Filter & Exposure & Brightness \\
   ~~(UT in 2008) &                      &        & (min) & (mag)$^{(a)}$ \\
   \noalign{\smallskip}
   \hline
   \noalign{\smallskip}
Sep 17 02:53--03:43 & IRSF/SIRIUS        & $JHK_S$    & 50.0  & 21.0$\pm$0.5 /
20.4$\pm$0.4 / 20.3$\pm$0.5 \\
Sep 17 07:57--09:39 & MPI/ESO 2.2m/GROND & $g'r'i'z'$ & 75.0  & $\!\!$$>$23.6 /
22.81$\pm$0.07 / 22.05$\pm$0.05 / 21.76$\pm$0.05\\
Sep 17 07:57--09:39 & MPI/ESO 2.2m/GROND & $JHK_S$    & 60.0  & 21.50 $\pm$
0.06 / 21.29 $\pm$ 0.08 / 21.10 $\pm$ 0.15\\
Sep 19 08:04--09:42 & MPI/ESO 2.2m/GROND & $g'r'i'z'$ & 79.4  & $>$ 23.6 / $>$
23.8 / 23.47 $\pm$ 0.13 / $>$ 23.8\\
Sep 19 08:04--09:42 & MPI/ESO 2.2m/GROND & $JHK_S$    & 64.0  & $>$ 21.9 / $>$
21.2 / $>$ 20.5\\
Sep 20 08:42--09:42 & MPI/ESO 2.2m/GROND & $g'r'i'z'$ & 50.0  & $>$ 23.9 / $>$
24.2 / 23.78 $\pm$ 0.16 / $>$ 23.8\\
Sep 20 08:42--09:42 & MPI/ESO 2.2m/GROND & $JHK_S$    & 40.0  & $>$ 22.5 / $>$
21.5 / $>$ 20.6\\
Sep 24 07:32--09:31 & MPI/ESO 2.2m/GROND & $g'r'i'z'$ & 100.1~~ & $>$ 25.0 / $>$
24.5 / $>$ 24.3 / $>$ 23.9\\
Sep 24 07:32--09:31 & MPI/ESO 2.2m/GROND & $JHK_S$    & 74.2  & $>$ 22.2 / $>$
21.5 / $>$ 20.8\\
Oct 29 07:59--08:31 & Gemini-S/GMOS      & $i'$       & 24.0  & $>$ 25.1 \\
  \noalign{\smallskip}
   \hline
  \noalign{\smallskip}
\end{tabular}

\noindent{
  $^{(a)}$ Not corrected for Galactic foreground reddening of E(B-V) = 0.32 mag
  \cite{sfd98}.
  All magnitudes are given in the AB system.}
\end{table*}

\begin{table*}[ht]
\caption{Local photometric standards within 2\amin\ of the GRB.
Calibration of the field in $JHK$ was performed using 2MASS stars. 
The magnitudes of the selected 2MASS stars were then transformed
into the GROND filter system and finally into
AB magnitudes using $J$(AB) = $J$(Vega) + 0.91, $H$(AB) = $H$(Vega) + 1.38,
$K$(AB) = $K$(Vega) + 1.81 
\cite{gbc08}. Systematic errors are $\pm$0.02 mag for $g'r'i'z'$,
and $\pm$0.05 mag for $JHK_s$. 
}
\tiny
\begin{tabular}{lcccccccc}
  \hline
  \noalign{\smallskip}
  $\!\!\!\!$No$\!\!\!\!$ & Coordinates (J2000) & $g'$ & $r'$ & $i'$ & $z'$ & $J$
 & $H$ & $K_S$ \\
  \noalign{\smallskip}
  \hline
  \noalign{\smallskip}
1 & 07:59:28.97 -56:38:24.0 &
    17.59$\pm$0.01 & 16.82$\pm$0.01 & 16.50$\pm$0.01 & 16.27$\pm$0.01 &
    16.24$\pm$0.01 & 15.98$\pm$0.01 & 16.36$\pm$0.01 \\
2 & 07:59:27.40 -56:40:10.1 &
    17.18$\pm$0.01 & 16.18$\pm$0.01 & 15.79$\pm$0.01 & 15.52$\pm$0.01 &
    15.37$\pm$0.01 & 15.00$\pm$0.01 & 15.38$\pm$0.01 \\
3 & 07:59:24.01 -56:37:08.0 &
    17.11$\pm$0.01 & 16.31$\pm$0.01 & 15.98$\pm$0.01 & 15.72$\pm$0.01 &
    15.67$\pm$0.01 & 15.40$\pm$0.01 & 15.77$\pm$0.01 \\
4 & 07:59:19.84 -56:39:25.3 &
    17.90$\pm$0.01 & 16.92$\pm$0.01 & 16.54$\pm$0.01 & 16.28$\pm$0.01 &
    16.15$\pm$0.01 & 15.84$\pm$0.01 & 16.20$\pm$0.01 \\
5 & 07:59:17.70 -56:37:41.9 &
    18.11$\pm$0.01 & 17.09$\pm$0.01 & 16.71$\pm$0.01 & 16.45$\pm$0.01 &
    16.36$\pm$0.01 & 16.02$\pm$0.01 & 16.44$\pm$0.01 \\
   \noalign{\smallskip}
   \hline
   \label{stand}
\end{tabular}
\end{table*}

 Long-duration  Gamma-Ray   Bursts  (GRBs)  are  the  high-energy
signatures of the  death of some massive stars and  emit the bulk of
their radiation in the 300--800\,keV band.
In a small number of  events, emission up to $\sim$100\,MeV  has been detected,
  e.g.,   with   SMM   \cite{has98},   COMPTEL   \cite{hkm05},   EGRET
  \cite{kgp08},   and  recently   with   AGILE  \cite{gmf08}.    These
  high-energy photons offer  unique access to the  physics of GRBs.
  Firstly, the shape of the spectrum provides direct information about
  the  gamma-ray  emission  mechanism  
  (Pe'er et al. 2007, Giannios 2008).
Secondly,
  high-energy  photons  can place  tight  constraints  on the  Lorentz
  factor of the ejecta via the pair-production threshold.  Furthermore, in
  some  cases, the origin  of the  high-energy component  differs from
  that of the low-energy  emission (e.g., GRB~941017; \cite{gdk03}) or the
high-energy photons arrive  with a significant time delay  (e.g., $>$ 1hr in
  GRB~940217; \cite{hdm94}). The formation  of these properties is far
  from understood and can only  be addressed with an increasing number
  of bursts with GeV detections. Finally, the search for
  signatures of absorption against the intergalactic UV background light
  using the shape of the high-energy spectrum as well as the search for
  quantum gravity dispersion effects over cosmic distances in the light curve
  are implications of much broader scientific interest \cite{accpap}.

  An important prerequisite to the interpretation of the GeV component
  of a  burst is the accurate  knowledge of the distance.  Only few of
  the   previously  detected  GRBs   with  high-energy   emission  had
  identified optical  afterglows, as the  localization capabilities of
  high-energy missions were insufficient to facilitate rapid
  follow-up observations.

The recently  launched \emph{Fermi} Gamma-Ray Space  Telescope has the
ability to localise high energy  events using the Large Area Telescope
(LAT)
and  to measure spectra over a  large energy range
in  combination  with the  Gamma-Ray  Monitor  (GBM)  (8\,keV to
  300\,GeV).  The  \emph{Swift} satellite \cite{gcg04}  can also slew
rapidly to LAT locations and provide positions with arcsec-accuracy by
the detection  of the  X-ray afterglow, facilitating  and dramatically
enhancing the likelihood  of a distance measurement.

 The bright GRB~080916C was detected  by the GBM on 2008 Sep 16th,
  at 00:12:45 UT  \cite{gvh08}. The burst was located  in in the field
  of view of the LAT and emission above 100\,MeV was quickly localized
  \cite{tbc08}.   Follow-up observations  with the  \emph{Swift} X-ray
  telescope (XRT)
provided an X-ray afterglow candidate
  \cite{ken08}  which subsequently led  to the  discovery of  a faint
  optical/NIR source with GROND  \cite{crg08a} and SIRIUS \cite{nag08}.
 Further monitoring in both
  X-rays \cite{str08} and in the optical \cite{crg08b} established the
  fading and confirmed the source to be the afterglow of GRB~080916C.

GRB~080916C was also detected by other satellites in addition to Fermi
\cite{hgg08}:  AGILE  (MCAL, SuperAGILE,  and  ACS), RHESSI,  INTEGRAL
(SPI-ACS),  Konus-Wind,  and   MESSENGER.   The preliminary analysis
of the  GBM  integrated
  spectrum over a duration (T$_{90}$) of  66\,s results in a best fit
 Band function
\cite{bmf93} with E$_{peak}$ = 424$\pm$24\,keV, a low-energy photon index
$\alpha$ = --0.91$\pm$0.02, and  a high-energy index $\beta$ = --2.08$\pm$0.06,
giving a fluence    of
1.9$\times$10$^{-4}$\,erg  cm$^{-2}$ in  the 8\,keV  --  30\,MeV range
\cite{vhg08}.  The spectral results reported by RHESSI and
Konus-Wind \cite{gam08} are in broad  agreement.

\begin{figure}[ht]
\centering
\includegraphics[width=8.8cm]{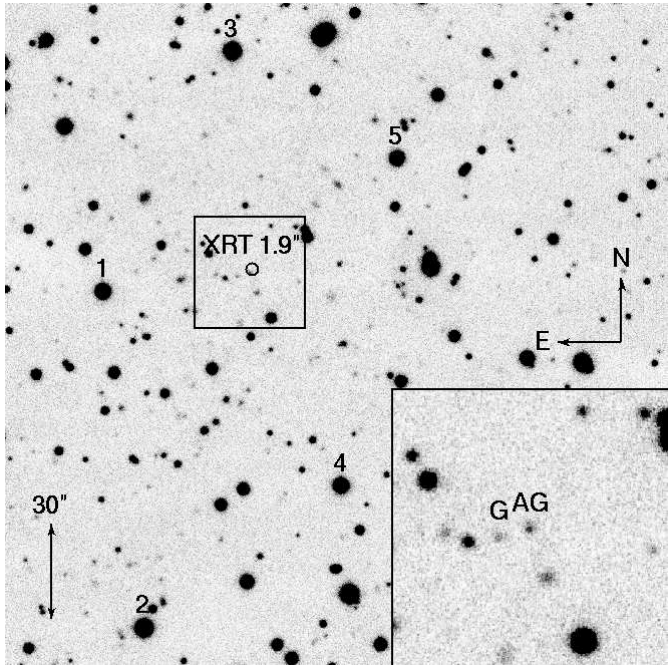}
   \caption[fc]{$i'$-band image
     of the afterglow of GRB~080916C obtained with the
     7-channel imager GROND at the 2.2m
    telescope on La Silla / Chile 32\,hrs after the burst.
    The circle denotes the {\emph Swift}/XRT error box.
    Some of the local standard stars of Tab. \ref{stand} are labeled. 
    A zoom into the innermost region is shown in the bottom-right,
    with the afterglow (AG) and a galaxy (G) 4$^{\prime\prime}$ 
    from the afterglow labeled.}
    \label{fc}
\end{figure}

The measurements of the high-energy emission from GRB 080916C by the
instruments of the Fermi Gamma-Ray Space telescope are described 
in Abdo et al. (2009). Here we report on the discovery of the optical/NIR
afterglow of GRB 080916C, the measurement of its redshift, and consequently
on the recognition of its extreme explosion energy and the large Lorentz factor
of its relativistic outflow.

\subsection*{The GRB afterglow}

\begin{figure}
\includegraphics[width=8.8cm]{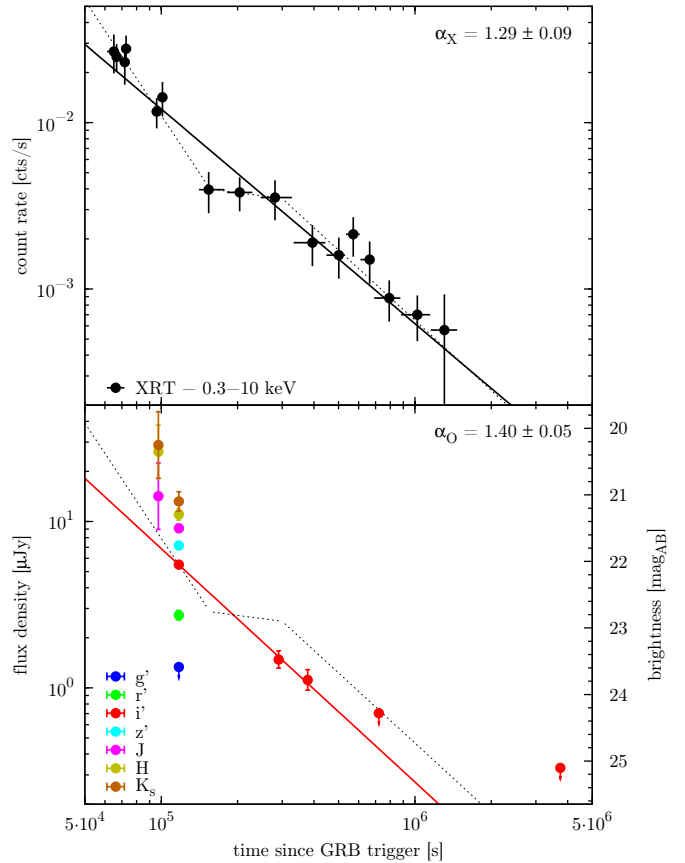}
\caption[lc]{X-ray (upper panel) and optical/NIR light curve  (lower panel) 
  of the GRB~080916C afterglow.  The solid lines mark the best fit
  power laws to the X-ray and $i'$-band data (labeled in the upper right
  corner). The power law segments as given by  Swift \cite{str08} are
  shown as dashed lines; they were scaled to the $i'$-band in the lower
  panel to show that this 3-segment power law does not fit the optical data,
  i.e. the apparent X-ray plateau phase 
  is not  detected in  the optical  data. 
  However,  the X-ray light  curve can also be fit with  a single
  power law, with only marginally larger $\chi^2_{red}$, as compared to
 the 3-segment  powerlaw, which results in  $\alpha_X$ = 1.29$\pm$0.09,
   The optical decay is mainly constrained by the  three $i'$-band data 
  points. 
  The $JHK_S$ observations of SIRIUS and GROND during the first night  are 
  consistent with this decay.}  
    \label{olc}
\end{figure}

A comparison  of GROND observations from Sep. 17 and 19, 2008 clearly
reveals a  fading  source inside the  {\emph Swift}/XRT error box 
(Fig.  \ref{fc}), with coordinates RA (J2000.0)
=  07\h 59\m  23\fss32, Decl.  (J2000.0) =  --56\degs 38\amin18\farcs0
(0\farcs5 error).  The  decay between 1.3 to 4\,d after the GRB
is  well   described  by  a   single  power  law  with   $\alpha_O$  =
1.40$\pm$0.05 (Fig.  \ref{olc}),
compatible within the errors to the X-ray decay slope 
$\alpha_X$ = 1.29$\pm$0.09.

\begin{figure}
\includegraphics[angle=270, width=1.0\columnwidth]{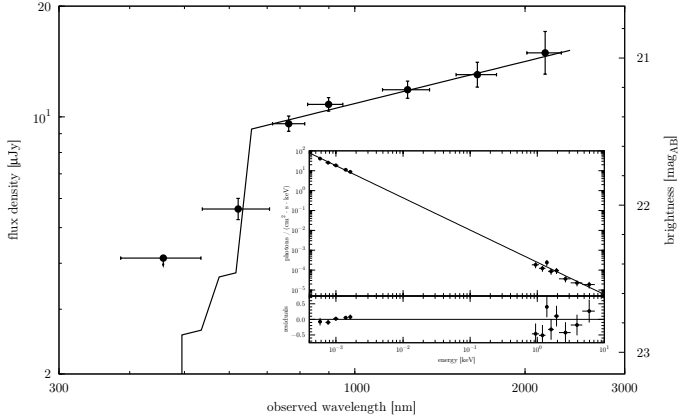}
\caption[sed]{Spectral  energy  distribution of  the  afterglow,
    derived from the coaddition of  all GROND exposures from the first
    night (Sep.   17, 2008). The SED is  best fit with a  power law of
    spectral   index  $\beta=0.38$,  no  extinction,   and  Ly$\alpha$
    absorption at  a redshift of  $z=4.35\pm0.15$.  Alternatively, the
    $r'$-band  drop  could be  attributed  to  reddening  in the  host
    galaxy,  caused either by  substantial UV  absorption or  a strong
    broad absorption  feature like that at 2175 \AA\ \cite{kkg08}. 
    However, the resulting host extinction
    corrected  spectral slope  of  $\beta$ \lax\  0.0  would  be
    incompatible with  most theoretical  models \cite{spn98} and with the
    X-ray spectrum.   The lack of  curvature in the $i'-K_s$  SED, and
    its extrapolation to the X-ray data, provides additional arguments
    against  host extinction.   Similarly,  a spectral  break can  not
    easily explain the $r'-i'$ color without redshift as the difference in
    power  law index  would  be  2.5, much  larger  than predicted  by
    theory\cite{spn98}.   In  addition,  the  steep  power  law  would
    significantly underpredict the observed X-ray fluxes.
    The inset shows that the best-fit GROND power law connects without
    break or offset to the Swift/XRT data, supporting the correct
    modelling of the GROND SED.
    }
    \label{sedag}
\end{figure}

\begin{figure}
\centering
\includegraphics[angle=0, width=1.05\columnwidth]{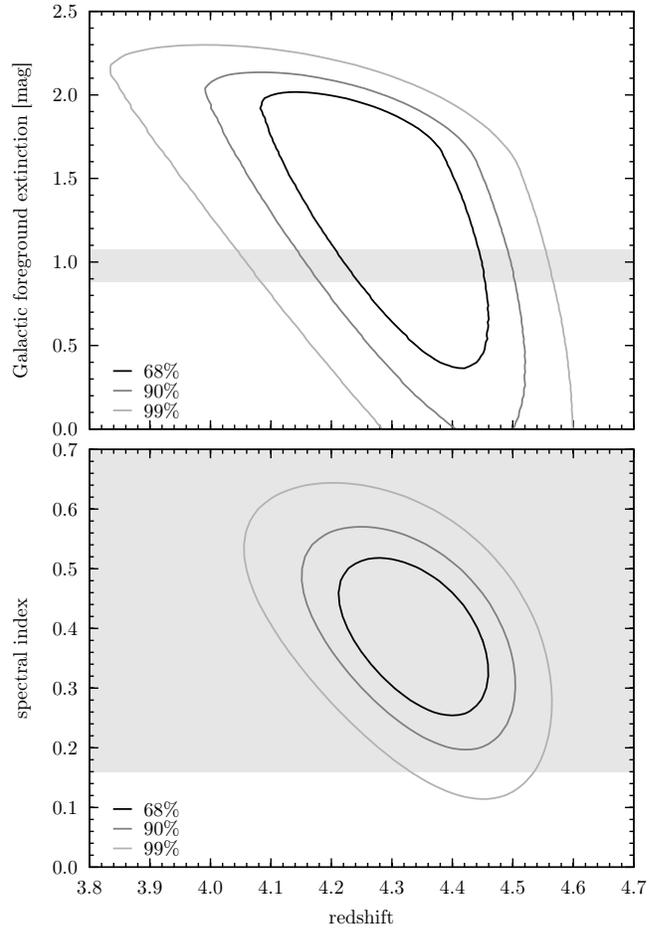}
\caption[chi]{Contour plot of the foreground A$_{\rm V}$ (upper panel)
     and spectral index (lower panel) from the spectral energy
     distribution fit against the redshift. This shows that the photo-z 
    determination is stable against uncertainties in the correction for
    the galactic foreground extinction, as well as the powerlaw index.
    The 68\%, 90\% and 99\% confidence contours are plotted.
    The shaded area in the upper panel shows the A$_{\rm V}$ range according to
    \cite{sfd98}. In the bottom panel the
    shaded area shows the power-law spectral index from
    the X-ray spectral fits.
    }
    \label{chi}
\end{figure}

A spectral  energy distribution (SED) was constructed  using the GROND
magnitudes  from  the first  night  of  observations (Tab. \ref{log}).   
The photometrically calibrated data (Tab. \ref{stand})  were
corrected for the  foreground galactic reddening of E(B-V)  = 0.32 mag
\cite{sfd98} corresponding  to an extinction  of $A_V$ = 0.98  mag and
fit by  an intrinsic power  law ($F_\nu \propto \nu^{-\beta}$)  plus three
different   dust  models,   as  well   as  without   extinction  (Tab.
\ref{SEDtable}).  The $i'$ to $K_S$ band data are best fit with a power
law slope of $\beta$  = 0.38$\pm$0.20  and  no host-intrinsic  extinction.
The  $i'$ to $r'$ band measurements deviate significantly and can be 
best explained with a Ly-$\alpha$ break at  $z$  =   4.35$\pm$0.15
(see  Tab.  \ref{SEDtable} and Fig. \ref{sedag}, and the rejected alternative
explanations given in the figure caption).  
The redshift  values resulting from all the fitted
models are  compatible, and the redshift error includes already the dependence
on the error of the photon index as well as foreground-extinction correction
(Fig. \ref{chi}). 
The $g'$-band upper limit  is consistent with
the  high-redshift  result,  though  it is  not  deep enough  to
constrain the fit.

\begin{table}
\caption{Results of the spectral energy distribution fitting without dust and 
with dust models from the Milky Way (MW),
Large Magellanic Cloud (LMC), Small  Magellanic Cloud (SMC). 
The redshift and $\beta$ errors  are at the 2$\sigma$ confidence level.}
\begin{tabular}{lcccc}
  \noalign{\smallskip}
  \hline
  \noalign{\smallskip}
  Dust model   & Redshift & $\beta$ & $A_V^{host}$ (mag) & $\chi^2_{red}$ \\
  \noalign{\smallskip}
  \hline
  \noalign{\smallskip}
  none & 4.35$^{+0.12}_{-0.13}$ & $0.38^{+0.20}_{-0.19}$ & --  & 1.04  \\
  MW   & 4.35$^{+0.12}_{-0.16}$ & $0.38^{+0.21}_{-0.23}$ & $0.0^{+0.4}_{-0.0}$ &1.04\\
  LMC  & 4.28$^{+0.15}_{-0.24}$ & $0.34^{+0.19}_{-0.24}$ & $0.1^{+0.4}_{-0.1}$ &0.95\\
  SMC  & 4.35$^{+0.13}_{-0.26}$ & $0.38^{+0.13}_{-0.28}$ & $0.0^{+0.2}_{-0.0}$ &1.04\\
   \noalign{\smallskip}
   \hline
\end{tabular}
\label{SEDtable}
\end{table}

No X-ray measurements of the  afterglow are available during the time of
the initial GROND  epoch.  However, we  can use  the SIRIUS
$JHK_S$-band  brightness  as  well  as  a  back-extrapolation  of  the
afterglow decay  slope to re-scale the  GROND SED to  the earlier time
when  XRT measurements are  available.  The  \emph{Swift}/XRT spectrum
from 61--102  ks post-burst has been  reported to be well  fit with an
absorbed   power   law  spectrum   with   photon   index  $\Gamma_{\rm X}$   =
2.1$^{+0.9}_{-0.7}$ and  a   column   density  of   $N_{\rm  H}$   =
3.7$^{+3.3}_{-1.1}$ $\times$  10$^{21}$ cm$^{-2}$ \cite{str08}.  Using
the result of no excess  extinction, we re-fit the X-ray spectrum with
the column density fixed to the galactic foreground value ($N_{\rm H}$
=   1.5  $\times$   10$^{21}$  cm$^{-2}$),   and  obtain   $\Gamma_{\rm X}$  =
1.49$^{+0.31}_{-0.34}$,  consistent with  the slope  of the  GROND SED
(note that the  spectral index $\beta$ is related  to the photon index
$\Gamma$ from the  X-ray spectral fitting by $\beta_X$  = $\Gamma_{\rm X}$ - 1).
The GROND and  XRT combined SED is compatible with  a single power law
over the complete spectral range (see inset of Fig. \ref{sedag}).

No counterpart or  host galaxy was detected seven weeks after the
burst at the  position of the optical/NIR afterglow,  with a two-sigma
upper limit of $i' > 25.1$\,mag within a 1.1\asec-radius aperture
centered at the afterglow position.   This is not surprising given the
brightness distribution  of the  known GRB host  galaxies \cite{sgb08};
the brightness limit places a loose lower limit on the redshift of
$z>1$. 

\begin{figure*}[ht]
\centering
\includegraphics[width=17.5cm]{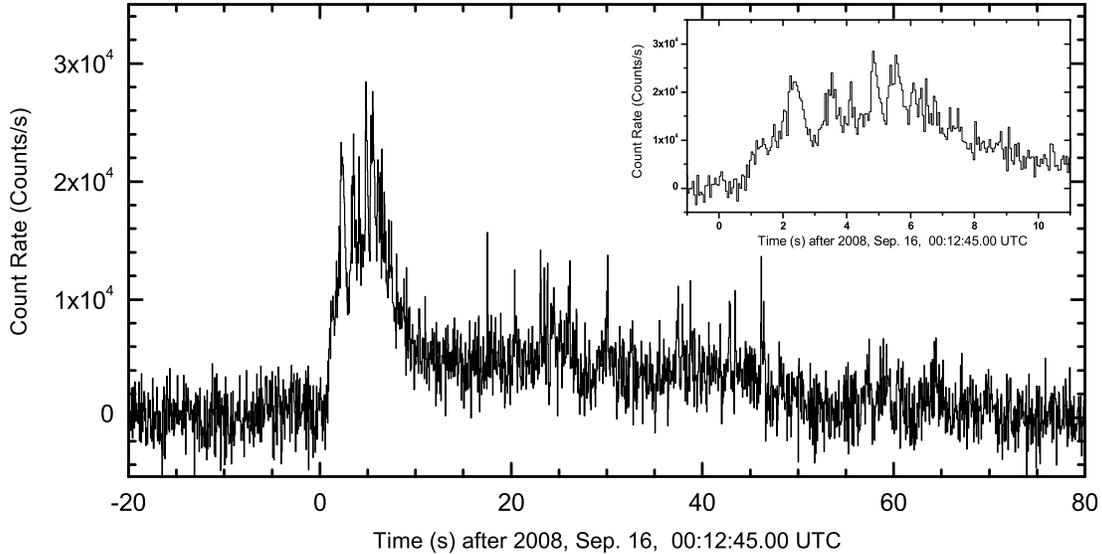}
\caption[acs]{Prompt emission light curve in the 80\,keV--30\,MeV
    energy band as measured with INTEGRAL SPI-ACS at 50 ms resolution.
    The inset shows a zoom of the main peak. Variability on time scale
    as short as 100\,ms is visible -- we measure several
    6$\sigma$ flux variations relative to the neighbouring data bins 
    on the 100 ms time scale.
}  
    \label{acs}
\end{figure*}

We note that the nearest object visible on our images is a galaxy
at 4\asec\ distance to the East. 
We obtained an optical spectrum of this galaxy beginning at
05:00 UT on 7 November 2008, using the Gemini-South telescope +
GMOS-S.  We obtained two spectra of 900\,s each
with the R400 grating centered
at 8000\,\AA, a second-order blocking filter in place, and with the
slit oriented to provide simultaneous coverage of the afterglow position.
Observations were carried out at relatively high airmass, with
1.7\asec\ seeing and variable sky background, and consist of two
spectra of 900\,s each.  Bias subtraction, flat-fielding, and
wavelength calibration were performed using the GMOS reduction package
under the {\it IRAF} environment.  
The wavelength solution was derived using a CuAr lamp spectrum taken
immediately after the science spectra.
Following extraction, the two one-dimensional spectra were coadded to
increase signal-to-noise, achieving a median S/N of 15.8 over the
6000--10,000\,\AA\ wavelength range.
We clearly detect the continuum of the resolved galaxy at wavelengths
greater than 6100\,\AA, placing an upper limit on the redshift of this galaxy
of $z < 4.0$, based on the absence of Lyman-alpha
absorption.  This is below the 99\% confidence range for the redshift
of GRB 080916C (Fig. \ref{chi}), and thus this galaxy is not related this burst.

\section*{Discussion and Conclusions}

It  is widely  believed that long GRBs are  produced in  the gravitational
collapse of a  massive star into a neutron star or  black hole. It has
been  argued  in the  past  that the  observed  total  energy in  GRBs
requires that  the emission is relativistic.  GRB~080916C is unrivaled
by all previous  events in this respect: with  its observed fluence in the  
8\,keV--30\,MeV band \cite{hgg08}, and redshift  of 4.35$\pm$0.15,  its total
isotropic energy    release   is    a    staggering   6.5$\times$10$^{54}$\,erg,
corresponding to 4 \msun$\times$c$^2$!

The  afterglow decay  slope is unlikely  a post-jet  break decay,
  which allows one to assume that the  jet break occurs  after the final
XRT observations  at 2$\times$10$^6$\,s. A break after this time 
would place a lower limit on the  jet half-opening  angle of
6\fdg1$\pm$0\fdg1  (ISM surrounding with  density of  1\,cm$^{-3}$) or
2\fdg2$\pm$0\fdg1 (wind  medium), which  in turn implies
lower limits for  the beaming-corrected energy release of $(3.7
\pm   0.1)    \times$10$^{52}$\,erg   (ISM)   or    $(4.9   \pm   0.1)
\times$10$^{51}$\,erg  (wind).  While  we  cannot distinguish  between
these two cases, we note that the energy for the ISM case is extremely
high,  exceeding  the previous  record  holders  GRB~990123 ($6  \times
10^{51}$\,erg) and 050904 ($7 \times 10^{51}$\,erg) 
by several factors. For the wind medium case, 
a jet break is expected to be very smooth, and 
might have started near the end of our coverage; thus the real
opening angle may not differ much from our lower limit. On the other hand,
using our measurements of the decay slope and spectral shape,
the cooling frequency would be still above the X-ray band at t $\sim$ 1 d,
which is unusual when compared to the majority of GRBs: thus the wind
interpretation is less likely than the ISM interpretation.

If the  emission were non-relativistic,  the required photon  field at
the  burst  location  would  be  optically thick  to  pair-production
(``compactness problem''; 
Ruderman 1975).   It has been recognized that
in  addition to  the  annihilation of  photons into  electron/position
pairs,  the  scattering of  photons  by  either  the electron  or  the
positron  created  in the  annihilation  process,  contributes to  the
optical  depth  of high-energy  photons  \cite{lis01}.  In fact,  this
latter  limit  is  in  many  cases more  constraining  than  the  pure
annihilation limit.

The most  sensitive instrument that detected  high-energy photons from
GRB  080916C   was   the  anti-coincidence  system (ACS)  of  the
spectrometer  onboard  INTEGRAL (SPI) \cite{rkl05}.   At  its  native  time
resolution SPI-ACS recorded  at peak more than 1200  counts per 50\,ms
in the  80\,keV--30\,MeV energy range.  This allowed  the detection of
variability  on   time  scales as short as 100 ms with high statistical
significance  (see   inset  of  Fig.~\ref{acs}). Using   eq.   9  of
Lithwick \& Sari (2001)
and the photon index of $\beta = -2.08$
as  measured  from GBM \cite{vhg08},  we
estimate a lower limit on the  Lorentz factor of the ejecta of $\Gamma
> 1090$.  The previously highest limit  on the Lorentz factor of a GRB
with  known redshift  using this  method  has been  $\Gamma >  410$
\cite{lis01}  for GRB  971214 at  $z=3.42$, for  which  the additional
assumption had to  be made that the photon  spectrum actually extended
to very high energies.  

Over the last two years, an alternative method
to determine the
initial Lorentz factor has been employed. This involves using
observations of the rising part of optical afterglows  to determine
when the blast wave has decelerated;
the  corresponding
Lorentz factor at the time of  the deceleration is expected to be half
of  the initial  Lorentz  factor $\Gamma_0$ \cite{sap99}.   
This method  provided
$\Gamma_0  \approx  400$  for  GRB  060418  and  060607  \cite{mol07},
$\Gamma_0 = 160$ for GRB  070802 \cite{kkg08}, $\Gamma_0 = 120$ for
GRB  080129 \cite{gre08}, and $\Gamma_0 = 200$ for GRB 071031
(Kr\"uhler et al. 2009).  Yet another method is based  on  the  evolution  
of the thermal  emission component in the  prompt emission of GRBs
\cite{prw07}, and also provides similarly low values  of  $\Gamma$.
It  is interesting to note
that our lower limit on $\Gamma$ for  GRB~080916C is  substantially larger  
than values determined by other methods.
Whether or not this is related to the GeV emission in GRB~080916C
remains to be seen.

The  high  signal-to-noise ratio of  the  SPI-ACS  data  is also  ideal  for
estimating the  variability of  the light curve,  a quantity  that has
been shown to correlate with the isotropic equivalent peak luminosity,
$L_{\rm iso,  peak}$. Following  the method described  in 
Li \& Paczy{\'n}ski (2006)
and using a smoothing time  scale of 13.7\,s, we derived a variability
index of $V=-2.26$ and a resulting $L_{\rm iso, peak}= (1.23 \pm 0.32)
\times 10^{52}$\,erg s$^{-1}$  (80\,keV--30\,MeV).  Using the observed
256\,ms peak  flux from Konus  \cite{gam08}, we derive  $L_{\rm iso,
  peak}  = 2 \times  10^{53}$\,erg s$^{-1}$  (20\,keV--10\,MeV).  

For the future, the synergy  of the detection of  GRBs with GeV  emission 
coupled with
the ability to localise and  determine redshifts for these events will
be extremely interesting as both $\Gamma$-determination methods can be
applied, providing a consistency  check of our picture of the GRB
and afterglow emission process.

\begin{acknowledgement}
TK acknowledges support by
the DFG cluster of excellence 'Origin and Structure of the Universe'.
Part of the funding for GROND (both hardware as well as personnel)
was generously granted from the Leibniz-Prize to Prof. G. Hasinger
(DFG grant HA 1850/28-1).
AvK acknowledges funding through DLR 50 QV 0301,
and XLZ through DLR 50 OG 0502.
This work made use of data supplied by the UK Swift Science Data Centre 
at the University of Leicester.
\end{acknowledgement}

\newpage


\begin{thebibliography}{}

\bibitem[(Abdo et al. 2009)]{accpap} Abdo A.A., Ackermann M., Ajello M., 
  et al. 2009, Science (in press)

\bibitem[(Band et al. 1993)]{bmf93} Band D., Matteson J., Ford L. et al. 1993, ApJ 413, 281

\bibitem[(Clemens et al. 2008a)]{crg08a} Clemens C., Rossi A., Greiner J.,  
et al. 2008a, GCN \#8257

\bibitem[(Clemens et al. 2008b)]{crg08b} Clemens C., Rossi A., Greiner J.,  
et al. 2008b, GCN \#8272

\bibitem[(Gehrels et al. 2004)]{gcg04} Gehrels N., Chincarini G., Giommi P., 
et al. 2004,    ApJ 621, 558

\bibitem[(Giannios 2008)]{gia08} Giannios, D., 2008, A\&A 488, L55

\bibitem[(Giuliani et al. 2008)]{gmf08} Giuliani, A., Mereghetti, S., Fornari, 
  F., et al. 2008, A\&A 491, L25

\bibitem[(Goldstein \& van der Horst 2008)]{gvh08} Goldstein A., van der Horst
 A., 2008, GCN \#8245

\bibitem[(Golentskii et al. 2008)]{gam08} Golenetskii S., Aptekar R., Mazets E.,
 et al. 2008,   GCN \#8258

\bibitem[(Gonzalez et al. 2003)]{gdk03} Gonzalez M.M., Dingus B.L., Kaneko Y.,
 et al. 2003,   Nat. 424, 749

\bibitem[(Greiner et al. 2008)]{gbc08}
Greiner J., Bornemann W., Clemens C., et al.
2008, PASP 120, 405

\bibitem[(Greiner et al. 2009)]{gre08} Greiner J., Kr\"uhler T., McBreen S., 
et al. 2009, ApJ (in press; arXiv:0811.4291)

\bibitem[(Harris \& Share 1998)]{has98} Harris, M.J.; Share, G.H. 1998, 
  ApJ 494, 724

\bibitem[(Hoover et al. 2005)]{hkm05} Hoover, A.S.; Kippen, R.M.; McConnell, 
  M.L. 2005, Il Nuovo Cimento C28, Issue 4, p.825

\bibitem[(Hurley et al. 1999)]{hdm94} Hurley K., Dingus B.L., Mukherjee R., 
 et al. 1994, Nat. 372, 652

\bibitem[(Hurley et al. 2008)]{hgg08} Hurley K., Goldsten J., Golenetskii S., 
et al.  2008,   GCN \#8251

\bibitem[(Kaneko et al. 2008)]{kgp08} Kaneko, Y.; Gonzalez, M.M.; Preece, R.D.; 
et al.  2008, ApJ 677, 1168

\bibitem[(Kennea 2008)]{ken08} Kennea J.A., 2008 , GCN \#8253

\bibitem[(Kr\"uhler et al. 2008)]{kkg08} Kr\"uhler T., K\"{u}pc\"{u} Yolda\c{s}
 A., Greiner J.,   et al. 2008, ApJ 685, 376

\bibitem[(Kr\"uhler et al. 2009)]{kgm09} Kr\"uhler T., Greiner J., McBreen S.,
  et al. 2009, ApJ (subm.)

\bibitem[(Li \& Paczy{\'n}ski 2006)]{lip06} Li, L.-X., Paczy{\'n}ski, B., 
2006, MN 366, 219

\bibitem[(Lithwick \& Sari 2001)]{lis01} Lithwick Y., Sari R., 2001, 
ApJ 555, 540

\bibitem[(Molinari et al. 2007)]{mol07}
  {Molinari} E., et al. 2007, A\&a 469, L13

\bibitem[(Nagayama et al. 2003)]{nag03}
Nagayama, T. et al.  2003, Proc. SPIE 4841, p. 459

\bibitem[(Nagayama 2008)]{nag08} Nagayama T., 2008 , GCN \#8274

\bibitem[(Pe'er et al. 2007)]{prw07} Pe'er, A.; Ryde, F.; Wijers, R.A.M.J.; 
et al. 2007,  ApJ 664, L1

\bibitem[(Rau et al. 2005)]{rkl05} Rau, A.; Kienlin, A. V.; Hurley, K.; 
Lichti, G. G. 2005,  A\&A 438, 1175

\bibitem[(Ruderman 1975)]{rud75} Ruderman M. 1975, Ann. NY Acad. Sci., 262, 164

\bibitem[(Sari et al. 1998)]{spn98} Sari, R., Piran, T., Narayan, R. 1998, 
ApJ 497, L17

\bibitem[(Sari \& Piran 1999)]{sap99} Sari R., Piran T., 1999, ApJ 520, 641

\bibitem[(Savaglio et al. 2009)]{sgb08} Savaglio, S.; Glazebrook, K.; 
Le Borgne, D. 2009, ApJ 691, 182

\bibitem[(Schlegel et al. 1998)]{sfd98} Schlegel, D., Finkbeiner, D., Davis, M. 
1998, ApJ 500, 525

\bibitem[(Stratta et al. 2008)]{str08} Stratta G., Perri M., Preger B. et al. 
2008,  GCN Report \#166.1

\bibitem[(Tajima et al. 2008)]{tbc08} Tajima H., Bregeon J., Chiang J., 
et al. 2008, GCN \#8246

\bibitem[(van der Horst \& Goldstein 2008)]{vhg08} van der Horst A., 
Goldstein A., 2008, GCN \#8278

\end{thebibliography}
\end{document}